%% file: main.tex
\documentclass[compsoc, conference, a4paper, 10pt, times]{IEEEtran}
\IEEEoverridecommandlockouts
\usepackage{cite}
\usepackage{amsmath,amssymb,amsfonts}
\usepackage{algorithmic}
\usepackage{graphicx}
\usepackage{textcomp}
\usepackage{bmpsize}
\usepackage{xcolor}
\usepackage{lipsum}

\usepackage{booktabs}

\usepackage[colorlinks=true,urlcolor=black]{hyperref}
\def\BibTeX{{\rm B\kern-.05em{\sc i\kern-.025em b}\kern-.08em
    T\kern-.1667em\lower.7ex\hbox{E}\kern-.125emX}}
\begin{document}

\title{Cyri: A Conversational AI-based Assistant for Supporting the Human User in Detecting and Responding to Phishing Attacks}

\author{\IEEEauthorblockN{1\textsuperscript{st} Antonio La Torre}
\IEEEauthorblockA{
\textit{Sapienza University of Rome}\\
Rome, Italy \\
latorre.2067686@studenti.uniroma1.it}
\and
\IEEEauthorblockN{2\textsuperscript{nd} Marco Angelini}
\IEEEauthorblockA{
\textit{Link Campus University of Rome)}\\
Rome, Italy \\
m.angelini@unilink.it}
}

\maketitle

\begin{abstract}
Phishing attacks have become increasingly sophisticated, exploiting human vulnerabilities through social engineering tactics to deceive individuals into revealing sensitive information. Traditional detection methods, such as blacklist-based and heuristic approaches, often fail to identify new or cleverly disguised phishing attempts due to their reliance on known patterns and technical indicators and not on the semantic characteristics of the attack attempt. This work introduces Cyri, an AI-powered conversational assistant designed to support a human user in detecting and analyzing phishing emails by leveraging Large Language Models (LLMs).\\
Cyri has been designed to scrutinize emails for semantic features used in phishing attacks, such as urgency, authority, impersonation, exclusivity, and undesirable consequences, using an approach that unifies features already established in the literature with others by Cyri features extraction methodology. Cyri can be directly plugged into a client mail or webmail, ensuring seamless integration with the user's email workflow while maintaining data privacy through local processing. By performing all analyses on the user's machine, Cyri eliminates the need to transmit sensitive email data over the internet, reducing security risks associated with external data breaches. 
The Cyri user interface has been designed to reduce habituation effects and enhance user engagement. It employs dynamic visual cues and context-specific explanations to keep users alert and informed while maintaining their experience in using emails.
Additionally, it allows users to explore identified malicious semantic features both through conversation with the agent and visual exploration, obtaining the advantages of both modalities for expert or non-expert users. It also allows users to keep track of the conversation, supports the user in solving additional questions on both computed features or new parts of the mail, and applies its detection on demand.\\
To evaluate Cyri's ability to distinguish between phishing and safe communications, we crafted a comprehensive dataset of 420 phishing emails and 420 legitimate emails. 
Through iterative evaluation, Cyri was optimized to reach an accuracy of 95.24\%, a precision of 96.8\%, a recall of 93.56\%, and an F1-score of 95.15\%, demonstrating high effectiveness in identifying critical phishing semantic features fundamental to phishing detection. A user study involving 10 participants, both experts and non-experts, evaluated Cyri's effectiveness and usability in real use, where the participants tested the system on their mail accounts. Results indicated that Cyri significantly aided users in identifying phishing emails and enhanced their understanding of phishing tactics.

\end{abstract}


\input{sections/01_introduction}

\input{sections/02_relatedwork}

\input{sections/03_design}
\input{sections/04_llm-model}
\input{sections/05_cyri-ui}
\input{sections/06_validation}
\input{sections/07_evaluation}
\input{sections/08_discussion}
\input{sections/09_conclusions}

\appendices
\section{Cyri Materials and Components}
All materials and source code of Cyri, including the created datasets for training, the video demonstration, and the results of different evaluation activities, are available at the following GitHub repository:\\\url{https://github.com/AntoReddy/Cyri}
\section{Topics Used for Phishing Email Generation and Example}
\begin{figure}[htbp]
  \centering
  \includegraphics[width=0.48\textwidth]{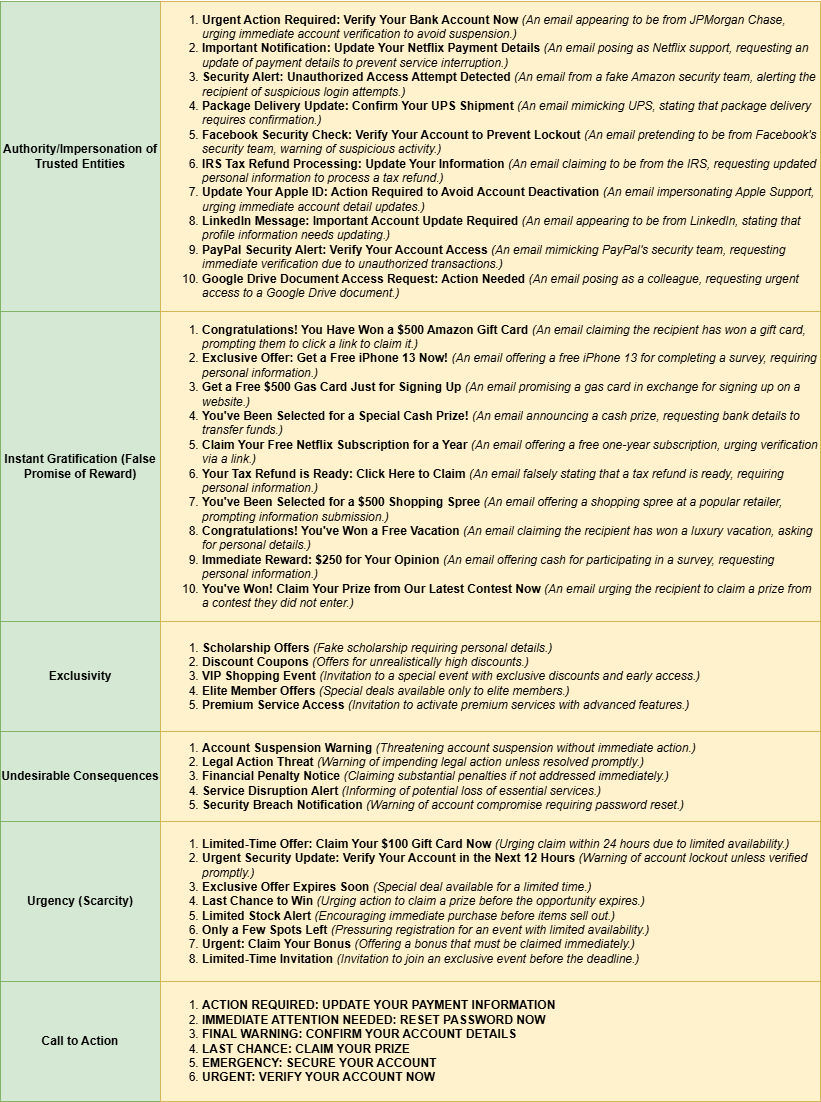}
  \caption{Topics for Generation of Phishing Emails, First Part}
  \label{fig:topicsgenerationemails}
\end{figure}

  We generated a total of 420 phishing emails, with 20 emails dedicated to each of
 the 21 identified semantic features. The emails were crafted to cover a wide array of topics relevant to each feature, enhancing the dataset’s heterogeneity.
 In the figure \ref{fig:topicsgenerationemails} are represented the topics utilized for the following features: Authority/Impersonation of Trusted Entities, Instant Gratification(False promise of reward), Exclusivity, Undesirable Consequences, Urgency (Scarcity) and Call to Action. 

\begin{figure}[htbp]
  \centering
  \includegraphics[width=0.48\textwidth]{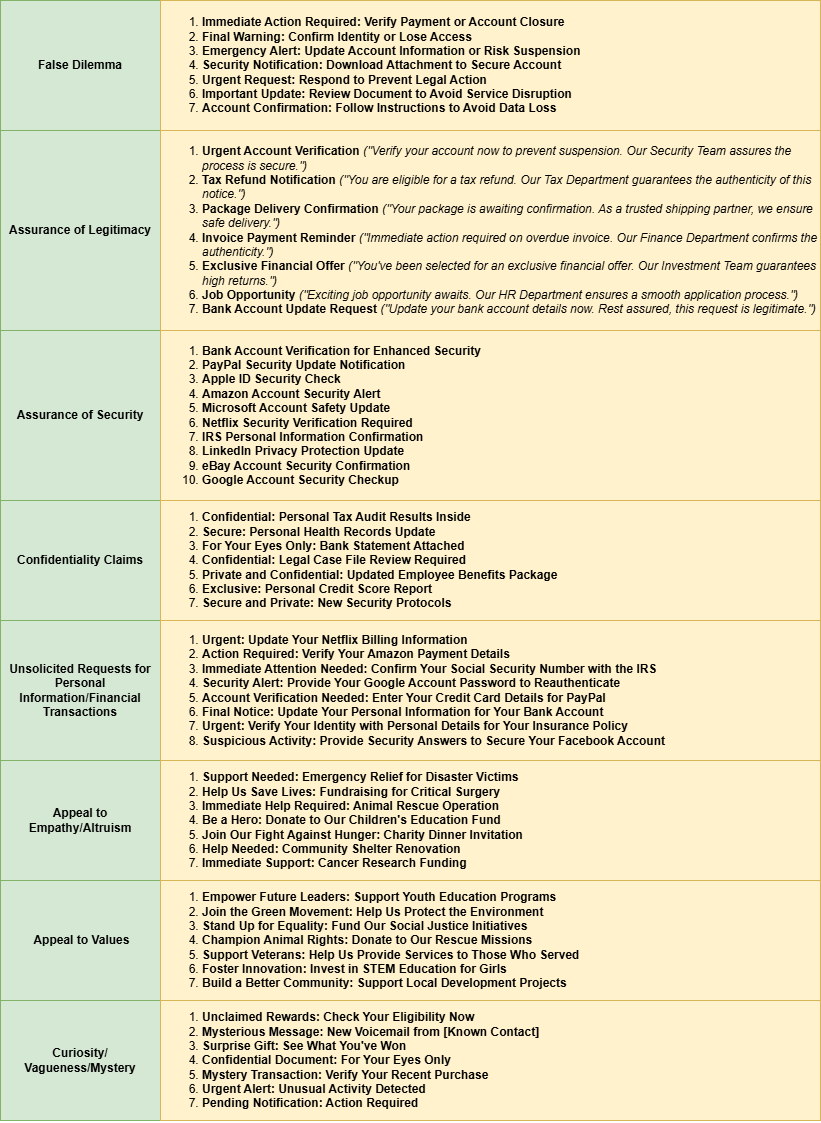}
  \caption{Topics for Generation of Phishing Emails, Second Part}
  \label{fig:topicsgenerationemails2}
\end{figure}

\begin{figure}[htbp]
  \centering
  \includegraphics[width=0.48\textwidth]{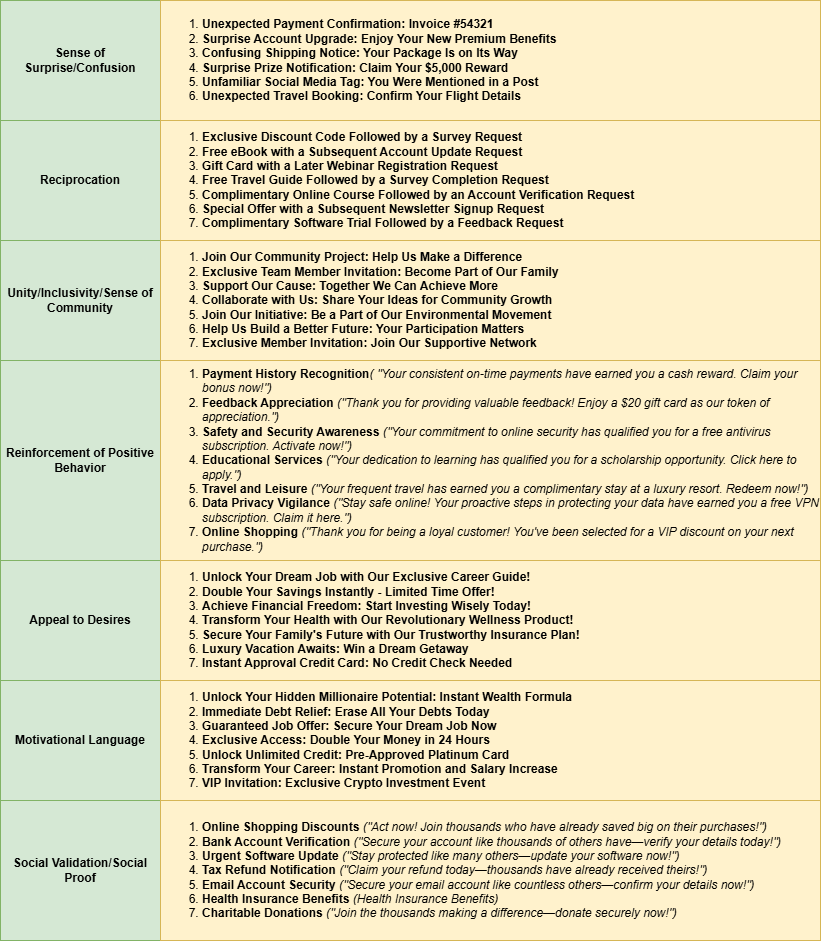}
  \caption{Topics for Generation of Phishing Emails, Third Part}
  \label{fig:topicsgenerationemails3}
\end{figure}

In the figure \ref{fig:topicsgenerationemails2} are represented the topics utilized for the following features:     False Dilemma,
    Assurance of Legitimacy,
    Assurance of Security,
    Confidentiality Claims,
    Unsolicited Requests for Personal Information/Financial Transactions,
    Appeal to Empathy/Altruism,
    Appeal to Values and
    Curiosity/Vagueness/Mystery.
In the figure \ref{fig:topicsgenerationemails3} are represented the topics utilized for the following features:     Sense of Surprise/Confusion,
    Reciprocation,
    Unity/Inclusivity/Sense of Community,
    Reinforcement of Positive Behavior,
    Appeal to Desires,
    Motivational Language and
    Social Validation/Social Proof.

\vspace{12pt}

\end{document}

%% file: sections/01_introduction.tex
\section{Introduction}
\label{sec:intro}
Among the various forms of cybercrime, phishing stands out as one of the most prevalent and harmful in modern time~\cite{b5}. According to the Anti-Phishing Working Group, over 1.2 million phishing attacks were recorded in the second quarter of 2023 alone~\cite{b6}. Every day, an estimated 3.4 billion phishing emails are sent by cybercriminals, amounting to over one trillion emails per year~\cite{b7}. Phishing serves as the entry point for 91\% of all cyber attacks and is involved in 36\% of all data breaches, making it the most common cause of such incidents. \\
Social engineering is at the core of phishing attacks, which exploits inherent human tendencies to trust and respond to certain stimuli~\cite{b1, b3, b19}. Attackers use psychological manipulation to deceive individuals into actions compromising security, such as clicking on malicious links or providing confidential information such as login credentials, financial details, or personal data~\cite{b2, b21}. This manipulation can have serious consequences, leading to identity theft, financial loss, and compromised personal or organizational security. \\
Human error is a significant factor in the success of phishing attacks, contributing to 95\% of successful cybersecurity breaches~\cite{b8}. Factors such as lack of awareness, inadequate training, stress, fatigue, and cognitive overload can impair an individual's ability to recognize and respond appropriately to phishing attempts~\cite{b9, b10, b17}. Attackers exploit these vulnerabilities by crafting messages that capitalize on distraction, curiosity, or the tendency to comply with authority figures~\cite{b4, b11, b20, b37}. This underlines the importance of implementing automated detection systems, which act as a first line of defense against human weaknesses. On the other hand, many of these solutions focus more on the technical identification of phishing without focusing on providing the human user explanations (apart from shallow ones or very technical ones, excluding non-technology expert people who represent the majority of targets) they can relate with, understand, and eventually use this information for improving awareness.\\
To better support non-expert human users in protecting against these threats, in this work, we propose Cyri, an AI-based solution designed to empower users in detecting, understanding, and responding to phishing emails. Cyri leverages Large Language Models (LLMs) to extract semantic features from email text, identifying subtle cues and psychological manipulation tactics that traditional detection systems might overlook or not target at all. By focusing on the meaning and context of the messages, Cyri provides users with an accurate assessment of potential threats at sentence-level granularity. Furthermore, users can clarify any doubts about the analysis by engaging in a conversation with Cyri, allowing them to gain deeper insights into why an email was flagged as phishing or safe. Cyri addresses the primary limitations of previous phishing detection solutions by integrating a user-centered interface that effectively communicates risks mitigating user desensitization from purely technical-based detectors~\cite{b22, b23}, while safeguarding user privacy through local data processing. These enhancements are crucial for improving phishing detection systems' overall effectiveness and acceptance in real-world applications.

Summarizing, the main contributions of this work are:
\begin{itemize}
    \item Instrumentation of local LLM to classify semantic features of email text linked to phishing attacks;
    \item A system, in the form of a plugin and web-based interface, which allows to directly connect Cyri to an existing email account without disturbing classic user experience while exploiting conversational capabilities through speech and interactive analysis through usable visual representations;
    \item Deep evaluation activities showing very good accuracy from the semantic features classifier and a user study with 10 participants, equally split into expert and non-expert, which tested Cyri directly on their email accounts and reported on the efficacy and usability of the proposed solution.
\end{itemize}

%% file: sections/02_relatedwork.tex
\section{Related Work}
\label{sec:rel}
Phishing detection is a critical area of cybersecurity research due to the increasing sophistication of phishing attacks and their significant impact on individuals and organizations.\\
One of the earliest and most straightforward methods for phishing detection involves the use of blacklists-based and rule-based approaches~\cite{b2}. The limitations of these traditional phishing detection methods are that they depend on known patterns, signatures, or previously identified malicious entities, making them inadequate for detecting new or evolving threats that do not match existing criteria \cite{b13}.
To overcome these limitations, researchers have increasingly turned to Machine Learning (ML)~\cite{b12, b15, b16, b18} and, more recently, Deep Learning (DL)~\cite{b14, b40} techniques.\\
Machine learning methods involve algorithms that classify emails, URLs, or websites as phishing or legitimate based on extracted features. Deep Learning Approaches involves training neural networks capable of automatic feature extraction from raw and unstructured data. These approaches aim to learn patterns from data, enabling the detection of previously unseen phishing attempts by generalizing from known examples.\\
Although many ML and DL approaches achieve strong performance, they primarily provide mostly statistical detection outcomes without explaining the reasoning behind them in a human-comprehensible form and focus on technical characteristics of the phishing attempts, struggling to identify the semantic and contextual subtleties of more complex phishing attempts~\cite{b24}. Understanding the reasoning behind their predictions is difficult, which can hinder trust and adoption, especially in critical security applications. The lack of transparency poses challenges in validating model decisions, diagnosing errors, and complying with regulatory requirements that may mandate explainability in automated decision-making systems.
Finally, in all these cases, the provided explanation is targeted at only expert users, neglecting most of the technology-unsavvy population, as Cyri does.

Very recently, Large Language Models (LLMs) have emerged as powerful tools in the field of Natural Language Processing (NLP)  due to their remarkable ability to comprehend and generate human-like text~\cite{b38}. LLMs can comprehend context, summarize complex texts, answer questions, and engage in conversations that are contextually appropriate. Researchers have begun to explore how LLMs can enhance the identification of phishing emails by leveraging their advanced natural language processing abilities.
While these models offer numerous benefits in various fields, several studies highlight that as LLMs become more accessible, cybercriminals can produce high-quality phishing content at scale~\cite{b24, b25, b26, b35} \c. This implies that incorporating LLMs into detection strategies is crucial to keep pace with the evolving threats. In particular, their advanced language understanding could enable the detection of sophisticated phishing attempts that traditional methods might miss.\\
Greco et al.~\cite{b24} address the growing concern that cybercriminals can efficiently leverage LLMs to craft more convincing phishing emails. The primary objective of Greco et al.’s study was to evaluate the effectiveness of traditional ML models in detecting phishing emails crafted by LLMs. By examining whether ML classifiers can distinguish between human-written and LLM-generated phishing emails, the researchers aimed to understand the limitations of current detection approaches and highlight the need for advanced solutions to address this emerging threat, justifying the study of solutions like Cyri. Greco et al. found that while the ML models could achieve moderate accuracy, they struggled to distinguish between human-generated and LLM-generated phishing emails reliably. The research underscores the urgent need to explore new approaches, possibly involving LLMs themselves, to detect sophisticated phishing emails generated by AI.
Moreover, it highlights the need to incorporate human-centric solutions, which means enhancing user interfaces and warnings to better engage users and support them in recognizing phishing attempts, as Cyri does. \\ 
Li et al. approach~\cite{b28} involves constructing a multimodal knowledge graph that includes textual and visual information about legitimate entities. The LLM is used
to extract and interpret references within the email content, while the knowledge graph provides a factual basis for verifying the authenticity of these references. However, maintaining an up-to-date knowledge graph of legitimate entities, especially across diverse industries, would be resource-intensive. Additionally, the proposed study does not provide a system that can be implemented for use in real conditions.
Koide et al. proposed ChatSpamDetector~\cite{b27}, which utilizes LLMs like ChatGPT to analyze email content and provide detailed reasoning for its classifications. However, sending sensitive email data to third-party servers for analysis could raise significant privacy concerns.\\
The most similar contribution to Cyri is a very recent preprint (October 2024) by Desolda et al.~\cite{b39}, which proposes a tool based on OpenAI's GPT-4o to detect phishing emails and generate explanation messages to users about why a specific email is dangerous. While sharing with Cyri the main goal of supporting the human user in managing phishing, there are some key differences with our contribution: first, it focuses on chatGPT-based models, which can be accessed only online, raising privacy concerns when handling sensitive data such as email content. Second, the explanations generated are general and more concerned with the communication levels (through the four levels of warning provided) than with a systematic analysis and detection of phishing features that can be explored and analyzed directly in the email text by expert and non-expert users. Finally, they do not provide a fully local system that is easy to incorporate into existing email clients and accounts like Cyri does.
\\
Despite the advancements in leveraging Large Language Models (LLMs) for phishing detection \cite{b27, b28}, existing solutions exhibit significant limitations related to user engagement and data privacy. Firstly, many of these approaches do not adequately consider the user experience, relying heavily on traditional warning dialogs to alert users to potential threats. Studies have consistently demonstrated that such warning dialogs are often ineffective in alerting users to phishing threats due to two primary issues: a lack of user understanding of the risks involved~\cite{b22} and the phenomenon known as the habituation effect~\cite{b23}. The habituation effect occurs when users are repeatedly exposed to the same visual stimulus, such as generic phishing warnings, leading them to gradually ignore these alerts over time. This desensitization diminishes the effectiveness of security measures, as users may overlook critical warnings and fail to take appropriate action.\\
Secondly, some solutions transmit personal and potentially confidential information over the internet, exposing users to risks associated with data breaches and unauthorized access. 

To address these challenges, we introduce Cyri, an AI-powered conversational assistant that prioritizes both user engagement and data privacy in its design. Cyri employs a user-centered interface that moves beyond generic warning dialogs by providing clear, contextualized explanations of potential phishing threats within the user's email communications. By offering detailed analyses and actionable advice, Cyri enhances user understanding of the risks and encourages proactive behavior in managing suspicious emails. This approach mitigates the habituation effect by engaging users with dynamic and informative content, thereby maintaining their attention and responsiveness to security alerts. 

%% file: sections/03_design.tex
\section{Cyri Architectural Design}
\label{sec:design}
Cyri represents an innovative AI-powered conversational assistant designed to
help users detect and analyze phishing attacks within email communications. By leveraging a refined Large Language Model (LLM) through prompt
engineering and Chain of Thought techniques, Cyri provides users with detailed explanations of the suspicious features that could make an email potentially malicious, as well as the necessary countermeasures. Cyri’s architecture is composed of three main components:
\begin{enumerate}
\item LLM-based Interactive Semantic Analyzer component (LISA): it performs in-depth email analysis using a local Large Language Model (LLM) through APIs.
\item E-mail client plugin: it captures incoming emails and communicates with the
Electron application. A demonstrator is implemented for Thunderbird since It is an open-source email client that offers extensive customization capabilities through its support for add-ons~\cite{b32}.
\item Visual and Audio Conversational interface (VAC): it serves as the user interface, manages data storage, and allows a non-expert user to analyze the classification of e-mails, the main semantic reasons, and inquire more on it in both interactive visual and audio means. It is implemented as an Electron web application for generality and usability.
\end{enumerate}

\begin{figure}[htbp]
  \centering
  \includegraphics[width=0.48\textwidth]{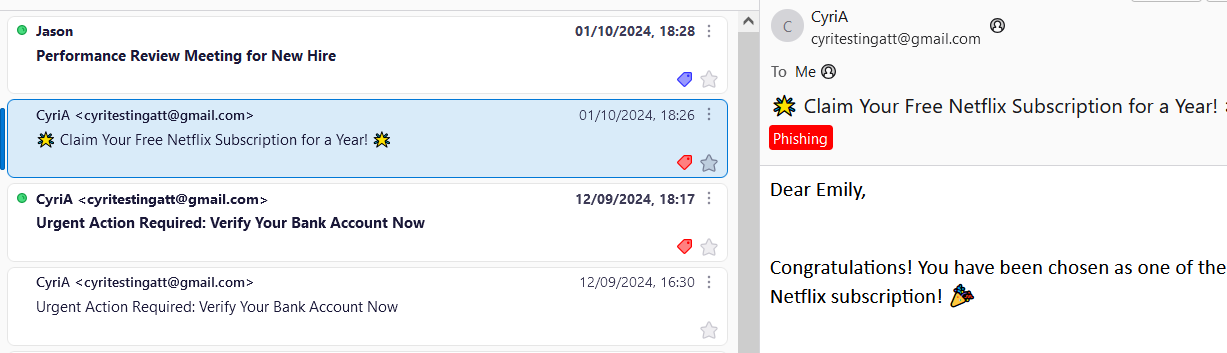}
  \caption{Cyri email Plugin Example using the Thunderbird email client}
  \label{fig:thunderbirdexample}
\end{figure}

Cyri continuously monitors incoming emails through the email client plugin. When a new, unseen email arrives, the plugin extracts essential data such as the sender's information, subject, body content, a flag indicating whether the sender is in the user's contacts, the message ID, and the timestamp. This data is then transmitted to both the VAC interface application and to the LISA component to perform an in-depth analysis of the email (locally hosted), specifically using the Meta-Llama-3.1-8B-Instruct model~\cite{b29}. LISA evaluates the email for semantic features such as urgency, authority, instant gratification, and others, all collected from the literature or extracted by LISA itself. LISA is helped by a sub-component for links checking that, using external APIs, specifically Google Safe Browsing~\cite{b30} and AbuseIPDB~\cite{b31}, enhance detection capabilities by checking only links and domains against known malicious entities and provides additional context to the semantic analysis.\\
Upon completion of the analysis, the results are stored in the VAC application and sent to the e-mail client plugin for e-mail text tagging and classification as ``Phishing'' or ``Safe'' (see Figure~\ref{fig:thunderbirdexample}). Finally, the user is presented with this information in the VAC interface and can explore it as explanations and converse with Cyri with a mix of visual cues and audio.

\begin{figure}[htbp]
  \centering
  \includegraphics[width=0.45\textwidth]{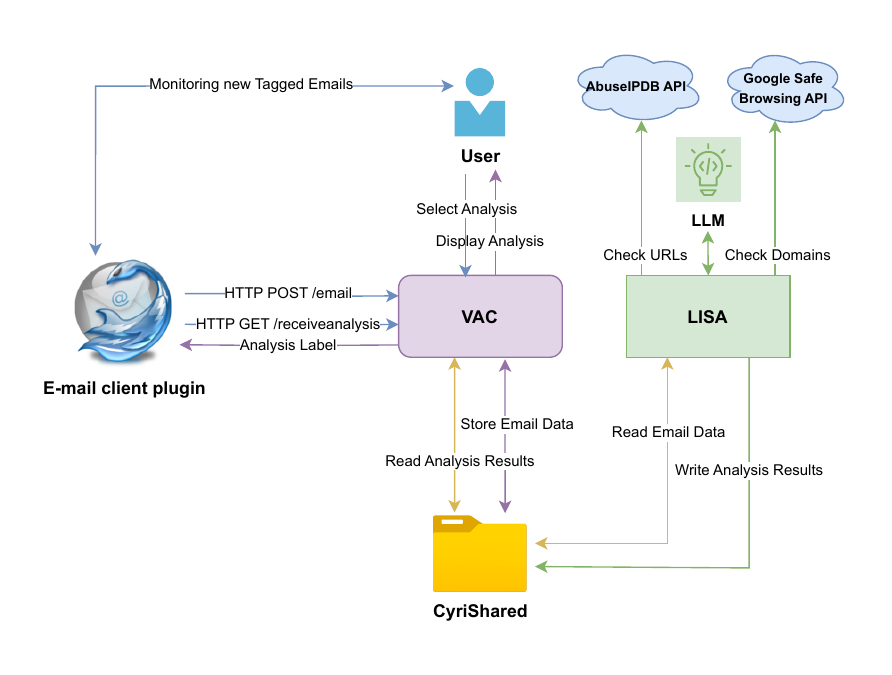}
  \caption{Cyri Architecture and Data Flow for Email Analysis}
  \label{fig:EmailAnalysisArchitecture}
\end{figure}

Figure~\ref{fig:EmailAnalysisArchitecture} illustrates the process by which Cyri analyzes the semantically tagged email in the VAC interface.

\begin{figure}[htbp]
  \centering
  \includegraphics[width=0.45\textwidth]{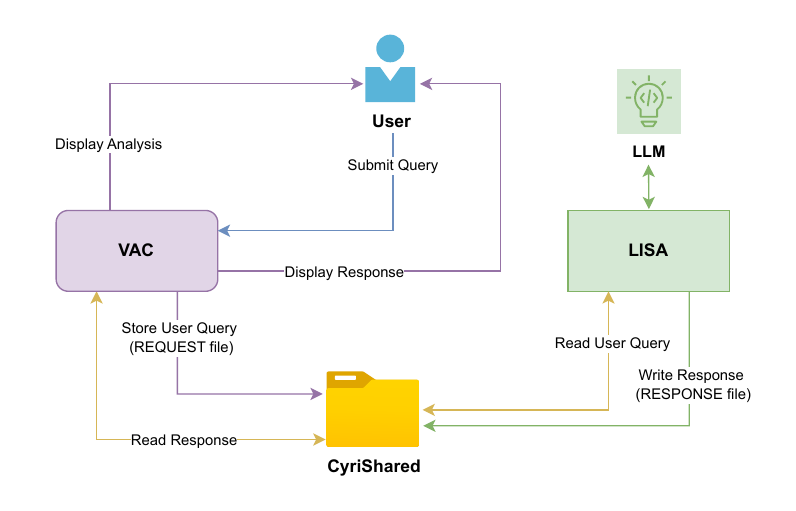}
  \caption{User Interaction and Query Processing Flow}
  \label{fig:ConversationArchitecture}
\end{figure}

The user can monitor newly tagged emails, interact with the detailed analysis through visual means, and issue further queries. User queries are processed interactively by the LISA component, which generates responses based on the conversation history with the user and initial semantic analysis of the e-mails, taking into account the user's inputs and questions. The whole process is visible in Figure~\ref{fig:ConversationArchitecture}.


Finally, security and privacy are integral to Cyri’s design, especially given the sensitive
nature of email content. Data privacy is ensured thanks to local processing and
minimal external data sharing (URLs and domains) for safety checks.
All email analyses are conducted locally on the user’s machine. Using a locally
hosted LLM ensures that sensitive information remains within the
user’s environment, mitigating the risk of data breaches. The LISA component implements the Hugging Face Transformers library to load and utilize the Llama 3.1 8B model locally.

%% file: sections/04_llm-model.tex
\section{Semantic Analysis of Phishing emails}
\label{sec:semantic}

Phishing attacks leverage sophisticated social engineering techniques to deceive recipients into disseminating sensitive information or performing actions compromising security. A critical aspect of enhancing phishing detection mechanisms involves
understanding and identifying the semantic features commonly employed in phishing emails. This section details the comprehensive collection of semantic features used and the instrumentation activities that guide the LISA component of Cyri in recognizing them into email text.

\subsection{Collection of Phishing Semantic Features}

This activity aims to create a robust dataset of phishing semantic features and emails containing them
that can inform the development of more effective detection algorithms and improve the capabilities of LLMs in identifying phishing
emails and recognizing the presence of these features in the text. The semantic phishing features collected in our dataset derive from two primary activities:

\begin{itemize}
    \item Literature-Identified Features review: it aims at collecting semantic features previously recognized and documented in academic and professional cybersecurity literature.
    \item Methodology-Extracted Features: it aims at extracting novel semantic features through a systematic extraction using an automatic text analyzer (i.e., ChatGPT-4) to each element of a comprehensive email phishing dataset created specifically for this purpose.
\end{itemize}

Table~\ref{tab:semantic-features} shows the results of these two activities, reporting the list of all the semantic features collected along with their corresponding source.

\begin{table}[ht]
\centering 
\caption{Overview of Cyri Semantic Features} 
\label{tab:semantic-features} 
\begin{tabular}{p{5.5cm}p{1.5cm}} 
\toprule 
\textbf{Semantic Feature} & \textbf{Source} \\ 
\midrule 
Authority &  \cite{b8, b20}\\
Impersonation of Trusted Entities &  Extracted\\ 
Instant Gratification &  \cite{b10}\\
Exclusivity &  Extracted \\
Undesirable Consequences & \cite{b10}\\ 
Urgency (Scarcity) &  \cite{b20, b37}\\
Call to Action &  Extracted\\
False Dilemma &  Extracted\\
Assurance of Legitimacy & Extracted \\
Assurance of Security &  Extracted\\
Confidentiality Claims &  Extracted\\
Unsolicited Requests for Personal Information & Extracted \\
Appeal to Empathy/Altruism & \cite{b10} \\
Appeal to Values &  Extracted\\
Curiosity/Vagueness/Mystery & Extracted \\
Sense of Surprise/Confusion & Extracted \\
Reciprocation &  \cite{b20, b37}\\
Unity/Inclusivity/Sense of Community & \cite{b20} \\
Reinforcement of Positive Behavior &  Extracted\\
Appeal to Desires & Extracted \\
Motivational Language &  Extracted\\
Social Validation/Social Proof &  \cite{b20, b37} \\
\bottomrule 
\end{tabular}
\end{table}

To identify and compile new semantic features, we first created a curated dataset of 300 phishing emails labeled by two experts in phishing analysis. This dataset was carefully assembled to include a
diverse range of phishing strategies and tactics. The sources from which  these emails were collected are:

\begin{itemize}
    \item Human-Generated Phishing Emails: Selected from the most recent ``Nazario'' and ``Nigerian Fraud'' collections~\cite{b33}, which are renowned repositories of real-world phishing emails that exhibit a variety of social engineering techniques;
    \item LLM-Generated Phishing Emails: collected by Greco et al.~\cite{b24}, which utilize advanced language models to generate realistic phishing emails that mimic human writing styles.
\end{itemize}

The dataset ensured comprehensive coverage of common and emerging phishing tactics by incorporating both human-generated and LLM-generated phishing emails.
Collecting the semantic features involved a meticulous analysis of each phishing email in the curated dataset being supported by experts and an automatic text analyzer (i.e., ChatGPT-4). Each email was input into the model with a carefully designed prompt that requested an in-depth examination of the email’s content, specifically focusing on the likelihood of it being a phishing attempt, the persuasion techniques employed, red flags, green flags, and potential countermeasures.
\\
From these responses, the identified persuasion techniques and red flags were first revised by experts and then documented. A validation process was undertaken to assess the significance and applicability of features not previously identified explicitly in the reviewed literature. This involved evaluating the consistency of these features across different phishing emails (validating their significance) and their effectiveness in deceiving recipients (validating their threat behavior). The validated features were then incorporated into the collection of semantic features, enhancing its coverage and utility. 

The semantic dataset provided to the Cyri LLM model comprises the feature names, an extensive description, and various examples. Table~\ref{tab:semantic-features-description} concisely describes each semantic feature we have identified.
\begin{table*}[ht]
\centering
\caption{Cyri Semantic Features Description}
\label{tab:semantic-features-description}
\begin{tabular}{|p{4cm}|p{12cm}|} 
\hline
\textbf{Semantic Feature} & \textbf{Description} \\ 

\hline
Authority & Impersonating authority figures to pressure recipients into complying with requests \\ [0.4em]
\hline
Impersonation of Trusted Entities & Mimicking trusted organizations deceives recipients into believing the email is genuine \\ [0.4em]
\hline
Instant Gratification & Offering tempting rewards prompts impulsive actions, exploiting the desire for quick benefits \\ [0.4em]
\hline
Exclusivity & Making recipients feel part of a select group increases compliance to avoid missing out on exclusive opportunities \\ [0.4em]
\hline
Undesirable Consequences & Threatening negative outcomes (e.g., account suspension) induces fear-driven responses without verification \\[0.4em]
\hline
Urgency (Scarcity) & Creating a sense of urgency forces recipients to act quickly, bypassing critical examination \\ [0.4em]
\hline
Call to Action & Clearly directing the recipient to perform a specific task (e.g., ``Click on the button below to verify your account'') \\[0.4em]
\hline
False Dilemma & Presenting only extreme choices pushes recipients toward the attacker’s desired course of action \\ [0.4em]
\hline
Assurance of Legitimacy & Convincing language and claims of authenticity are used to build trust and reduce suspicion \\ [0.4em]
\hline
Assurance of Security & Highlighting privacy and security reassures recipients \\ [0.4em]
\hline
Confidentiality Claims & Emphasizing the confidential nature of the information makes recipients feel they need to act without seeking advice or verification from others \\ [0.4em]
\hline
Unsolicited Requests for Personal Information & Requests for personal or financial data without prior authorization or legitimate justification \\ [0.4em]
\hline
Appeal to Empathy/Altruism & Exploiting the recipient's desire to help others or fulfill moral obligations \\ [0.4em]
\hline
Appeal to Values & Aligning with the recipient’s values builds trust and increases compliance with the attacker's requests \\ [0.4em]
\hline
Curiosity/Vagueness/Mystery & Vague or intriguing details induce recipients into taking action to satisfy their curiosity \\ [0.4em]
\hline
Sense of Surprise/Confusion & Unexpected scenarios create confusion, leading to unverified actions from recipients \\ [0.4em]
\hline
Reciprocation & Offering a benefit or favor creates a sense of obligation, leading recipients to fulfill follow-up requests \\ [0.4em]
\hline
Unity/Inclusivity/Sense of Community & Encouraging a sense of belonging or shared purpose motivates recipients to act in line with community goals \\ [0.4em]
\hline
Reinforcement of Positive Behavior & Praising the recipient for good behavior, and offering a reward reduces suspicion and increases engagement \\ [0.4em]
\hline
Appeal to Desires & Targeting personal goals or aspirations increases the chances of recipients ignoring warning signs \\ [0.4em]
\hline
Motivational Language & Evoking strong emotional responses, typically centered around desires for success, wealth, or security \\[0.4em]
\hline
Social Validation/Social Proof & Highlighting that others have taken the same action creates a sense of trust \\ [0.4em]
\hline
\end{tabular}
\end{table*}

\subsection{LISA: LLM-based Interactive Semantic Feature Analyzer}
\label{sec:lisa}

Traditional detection methods often rely on cloud-based services, which may not be suitable due to privacy concerns and dependence on external infrastructure. Deploying large LLMs raises privacy and data security concerns, as it requires sending sensitive emails to third-party servers. Users and organizations may be reluctant to adopt a system that necessitates sharing sensitive email content with external entities. 

To address these challenges, we developed a Python background process that performs in-depth phishing analysis and user query processing using a locally hosted LLM, specifically the Llama 3.1 8B model.
\\
We chose the Llama 3.1 8B model for its efficient reasoning capabilities and ability to handle contexts of up to 128,000 tokens, which is crucial when analyzing lengthy or complex emails. Additionally, this version is optimized for local deployment, balancing performance with resource demands, making it ideal for running on local machines without relying on cloud services.
\\
However, smaller models may not match the language comprehension of larger ones, making them more reliant on well-designed comprehensive prompts. For this reason, we defined an extensive prompt containing a large set of semantic social engineering techniques to improve the model's ability to detect diverse phishing tactics.

The LISA component performs two primary functions: analyzing incoming emails and handling user queries based on the analysis results during the conversation with the user. Each function is defined by a different prompt.

\subsubsection{Email Analysis Prompt}
\label{sed:eap}

The Email Analysis Prompt is a carefully constructed set of instructions designed to guide the LLM in performing a thorough analysis of an email to determine whether it is phishing or safe. The prompt employs several prompt engineering techniques to ensure that the LLM produces accurate, consistent, and user-friendly outputs. Due to the complexity and length of the prompt, we followed a Chain-of-Thought approach to make it more effective. Moreover, dissecting the prompt into its individual components ensures a thorough understanding of each aspect:\\

\noindent \textbf{1. Role Assignment}: ``\textit{You are an email phishing detector and analyzer. Your task is to identify whether an email is phishing or safe, explain why, and provide a detailed explanation.}''

The prompt begins by explicitly defining the LLM’s role as an ``email phishing detector and analyzer''. This sets the context for the model, focusing its capabilities on a specific task. By assigning a clear role, the LLM becomes ready to approach the subsequent instructions with the appropriate mindset.\\

\noindent \textbf{2. Presentation of the Email Content}: ``\textit{I want you to analyze the following email which could be phishing or safe: \{email\} I want you to tell me if this email is safe or phishing.}''

The prompt introduces the subject and the body of the email to be analyzed. Directly instructing the model to determine if the email is safe or phishing sets a clear objective.\\

\noindent \textbf{3. Base Reasoning Before Feature Consideration}: ``\textit{Use your base reasoning first to identify if the email is safe or phishing before considering the specified features.}''

The prompt instructs the LLM to use its inherent reasoning capabilities before relying on predefined features. This ensures that the model’s general understanding and language comprehension are utilized initially, potentially capturing nuances that this particular feature-based analysis might miss. \\

\noindent \textbf{4. Additional Information for Analysis and Guiding Questions:}: ``\textit{Here is additional information regarding the email for your analysis: \\
        1: Sender Information: \{sender\_email\} \\
        2: Google Safe Browsing API Result: \{google\_safe\_browsing\_output\}.\\
        3: AbuseIPDB Result: \{abuse\_ipdb\_output\}. \\
        - Is the sender domain or any URL found in the email reported as unsafe?\\
        - Identify if there is any impersonation of a well-known brand by comparing the sender’s email address with the claimed organization in the email content. If spoofing is detected, explain the inconsistencies. For example, if the email claims to be from 'Amazon' but the domain is not related to Amazon, highlight the inconsistency.\\
        Interpret the Google Safe Browsing API results: If threats are found, include the details. If no threats are found, note that.
        Interpret the AbuseIPDB results: If the domain is flagged as malicious, include the confidence score. If the domain is not flagged, note that as well. **Specify whether the domain refers to the sender or a link present in the email.** \\
        The sender's email address (\{sender\_email\}) is \{isSafeOutput\}.
        }'' \\
\indent The prompt provides external data such as the sender’s email, and results from security APIs enrich the context. These questions and instructions direct the LLM’s attention to specific aspects of the email, ensuring a comprehensive analysis. The LLM is instructed to determine whether the sender's domain or any URLs included in the email are reported as unsafe by the external APIs. The LLM is asked to compare the sender's email address with the organization mentioned in the email content to identify any impersonation. If spoofing is detected, for instance, the email claims to be from a reputable company like ``Amazon'' but the sender's domain does not match Amazon's official domain, the LLM should highlight these inconsistencies.
The variable \{isSafeOutput\} is set to indicate that the sender is ``present in the recipient's contact list and is trusted by the recipient'' or ``not present in the recipient's contact list''. This ensures that the model considers the trust relationship between the sender and the recipient. If the sender is recognized and trusted (i.e., in the contact list), the model lowers the phishing risk assessment for that email. \\

\noindent \textbf{5. Definition of Phishing/Safe Emails and Examples}: ``\textit{Here's a clear distinction for your analysis: \\ \\**Phishing Email**: Phishing emails are malicious attempts to deceive recipients into providing sensitive information or performing harmful actions.
            \\
            **Safe Email**: Safe emails are legitimate communications which typically have the following characteristics: Clear and concise language; Recognizable Sender Information; Content is relevant to the recipient's context (e.g., work-related updates, newsletters, transaction confirmations); Safe Links and Attachments. It includes, but is not limited to: routine communications like meeting requests, project updates, or a legitimate promotional email (Marketing email) from a company or organization offering products or services and it may contain offers, discounts, or promotional content.
            \\ \\
            I will provide examples of safe and phishing emails.\\
             This is a safe email:\\ \\
            \{example\_safe1\}.\\ \\
             This is a safe email:\\ \\ 
            \{example\_safe2\}.\\ \\ 
             This is a safe email:\\ \\
            \{example\_safe3\}.\\ \\
             This is a phishing email:\\ \\
            \{example\_phishing\}.
            }''
            
\indent Clear definitions and examples of phishing and safe emails are provided to help the model distinguish accurately between them for its classification process. We have added more examples of safe emails to improve the model’s ability to correctly identify safe emails and reduce false positives. \\
            
\noindent \textbf{6. Output Format Specification}: ``\textit{In the first line of the output, I want you to always respond with 'This email is [Likelihood Category] phishing ([percentage]\%)' or 'This email is [Likelihood Category] safe ([percentage]\%)' where you combine whether the email is phishing or safe with the likelihood description. \\ \\  Use these thresholds to categorize the likelihood of phishing: \\ \\
          - $0\% < x < 20\%$: Unlikely to be phishing \\ 
          - $20\% < y < 60\%$: Possibly phishing \\
          - $60\% < z < 90\%$: Likely phishing \\
          - $u > 90\%$: Almost certainly phishing \\ \\
    Also, categorize the likelihood of the email being safe:\\ \\
          - $0\% < x < 20\%$: Unlikely to be safe \\
          - $20\% < y < 60\%$: Possibly safe \\
          - $60\% < z < 90\%$: Likely safe \\
          - $u > 90\%$: Almost certainly safe}''

The prompt specifies the exact format for the output first line which will be composed by a percentage of the email being safe or phishing. By providing thresholds for likelihood categories, it ensures consistency in the model’s assessments and facilitates quantifiable evaluations. \\

\noindent \textbf{7. Feature Identification and Analysis}: ``\textit{You have to find the following features: \{features\}}''

With this step of the prompt we pass to the model the entire Cyri dataset of phishing semantic features composed by the features name, an extensive description and various examples to allow the model to perform a comprehensive analysis. \\

\noindent \textbf{8. Exact Output Format Instructions}: ``\textit{I want the output EXACTLY like this: \\  \\
- 'This email is [Likelihood Category] phishing ([percentage]\%)' or 'This email is [Likelihood Category] safe ([percentage]\%)'\\ \\
- Detailed Explanation: Provide a thorough explanation suitable for non-experts of why this email is phishing or safe. Clearly state your base reasoning for the classification, if spoofing is detected and if the sender is in the contact list or not (and how this impacts your assessment). Include references to specific elements of the email, the features of the email, the results from the Google Safe Browsing API, and the AbuseIPDB check, making sure to address how each contributes to your final assessment. \\ \\
- 'List of features found': [feature1; feature2; ...] **only the features present in the list below** for phishing emails. If the email is safe, define characteristics that make it safe (do not include any of the features present in the list below if the email is safe). \\ \\
- 'Analysis': \textless name of the feature \textgreater: '\textless specific part of the email \textgreater'. \textless explanation of why this part is linked to the feature \textgreater. **Only elements contained in 'List of features found' must be included**. \\ \\
- Countermeasures: where you offer practical recommendations on how the recipient should handle this email. These recommendations should be based on the identified risks and features, guiding the recipient on what actions to take next (e.g., verifying the sender, avoiding clicking on links, reporting the email as phishing, etc.).
}'' \\
\indent The prompt provides an exact template for the output, reducing variability and ensuring that all necessary components are included. Having a structured content analysis allows us to enhance the user interface design of the Cyri VAC component since it is possible to personalize the style of every section of the LLM analysis.\\

\noindent \textbf{9. Communication Style Guidelines}: ``\textit{Ensure the explanation is written in a conversational tone that directly addresses the recipient, making the analysis feel personalized. \\
Speak directly to the recipient using 'you' and 'your' when explaining why the email might be phishing or safe. \\
Provide clear, user-friendly explanations that are easy for non-experts to understand, directly addressing the recipient.}'' \\
\indent These instructions shape the tone and accessibility of the output, ensuring that it is appropriate for users without technical expertise.\\

\noindent \textbf{10. Feature Names and Weights}: ``\textit{Remember to use the exact names of the features listed below: \\ \\ \{list\_features\_names\} \\ \\ Weights are assigned to each feature indicating their importance in the classification: \\ \\ Authority, Impersonation of Trusted Entities: 0.6;
    Instant Gratification (False promise of reward): 0.9;
    Exclusivity: 0.8;
    Undesirable Consequences: 0.9;
    Urgency (Scarcity): 0.9;
    Call to Action: 0.9;
    False Dilemma: 0.8;
    Assurance of Legitimacy: 0.1;
    Assurance of Security: 0.3;
    Confidentiality Claims: 0.2;
    Unsolicited Requests for Personal Information/Financial Transactions: 0.9;
    Appeal to Empathy/Altruism: 0.4;
    Appeal to Values: 0.3;
    Curiosity/Vagueness/Mystery: 0.3;
    Sense of Surprise/Confusion: 0.3;
    Reciprocation: 0.3;
    Unity/Inclusivity/Sense of Community: 0.3;
    Reinforcement of Positive Behavior: 0.2;
    Appeal to Desires: 0.3;
    Motivational Language: 0.5;
    Social Validation/Social Proof: 0.5;
}''\\
\indent By specifying exact feature names, the prompt ensures consistency in terminology. Assigning weights to phishing features reflects their significance in identifying malicious emails. The weighting system guides the LLM’s reasoning process, emphasizing critical indicators.
Features assigned higher weights (e.g., 0.9) are considered strong indicators of phishing: Urgency (Scarcity); Undesirable Consequences; Unsolicited Requests. Features with moderate weights (e.g., 0.5 to 0.8) contribute significantly but may require the presence of additional indicators. Finally, features assigned lower weights (e.g., 0.1 to 0.4) may not strongly indicate phishing independently, but they can contribute to the overall assessment when combined with other indicators.

\subsubsection{Conversation Prompt}
The Conversation Prompt is designed to enable LISA to engage interactively with the user, providing detailed and context-aware responses to user queries based on prior email analysis. The prompt leverages previous interactions and analysis results to maintain continuity and relevance:
``\textit{You are an AI trained to analyze emails and interact with users to clarify and explain issues related to email security in simple terms. Use the analysis provided and the conversation history to inform your responses.\\
Given the user's query: (\{last\_user\_query\}), please provide a detailed and specific response that can help the user understand the steps to improve email security. For context I will give you also the initial instructions given to the model about how to analyze the email: \\ \\ \{initial\_prompt\} \\ \\ Here you can find the detailed analysis of the email generated by the model: \\  \{analysis\} \\ \\ All past interactions (questions and AI responses) related to this email analysis: \{conversation\_history\} \\ \\ Output only the response to this query: \{last\_user\_query\}
}''

%% file: sections/05_cyri-ui.tex
\section{Cyri Usable Interface }
\label{sec:ui}
\begin{figure*}[htbp]
  \centering
  \includegraphics[width=0.85\linewidth]{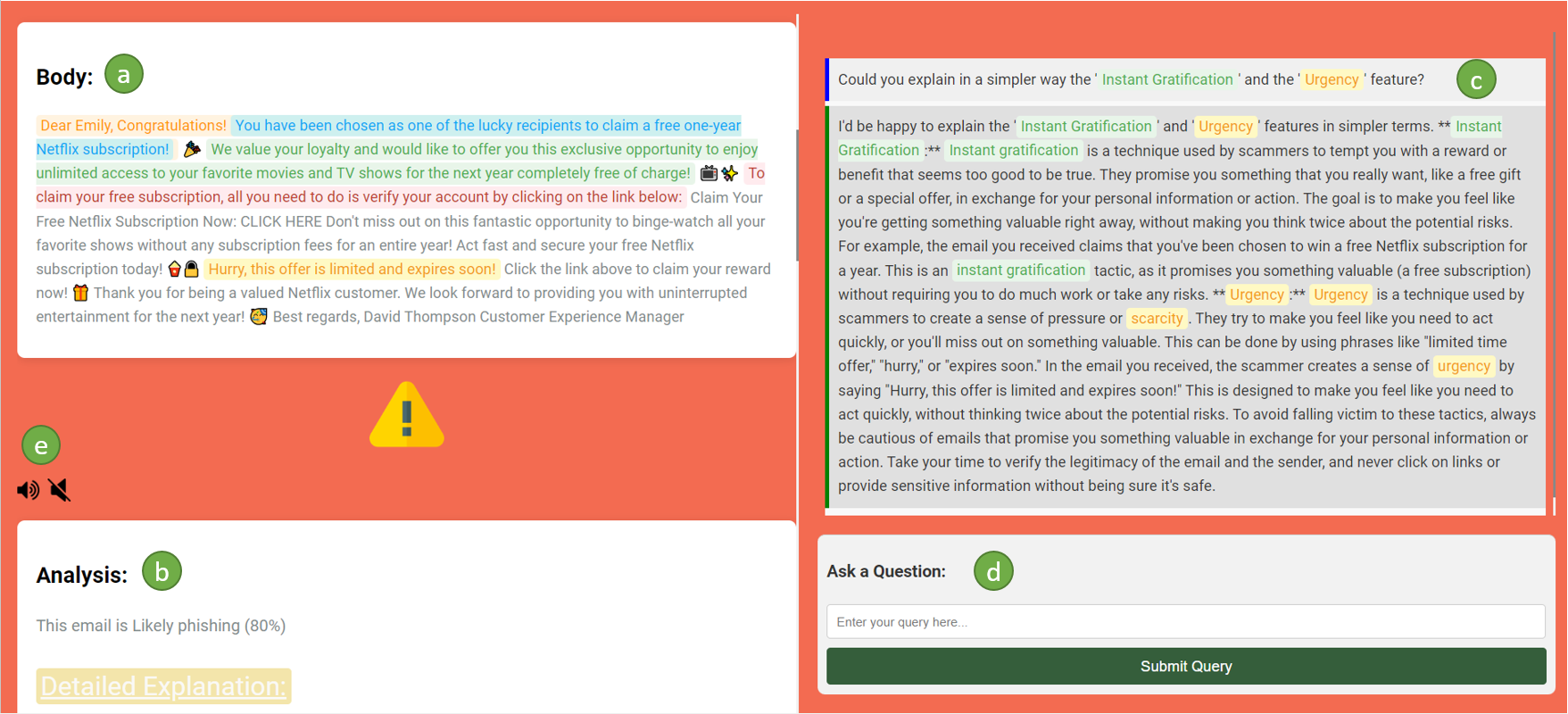}
  \caption{The VAC interface of Cyri in action: red background identifies a phishing mail, with semantic features highlighted in the email text (a) and the list below (b). Conversation with LISA happens on the right (c) through the query interface (d) or by audio (e)}
  \label{fig: examplePhishingInterface}
\end{figure*}

The VAC application serves as the central hub of Cyri, providing a user-friendly interface for email analysis and interaction. It is implemented using web technology and Electron. 
By leveraging Electron’s capabilities, the application offers a cross-platform, user-friendly interface that integrates seamlessly with LISA and the email client plugin.
Several studies~\cite{b22, b23, b36} have consistently demonstrated that traditional warning dialogs are often ineffective in alerting users to phishing threats due to a lack of user understanding and the habituation effect. The habituation effect occurs when users become desensitized to repetitive visual stimuli, such as generic phishing warnings, leading them to ignore these alerts over time and diminishing their vigilance. This desensitization results in users dismissing important security warnings without adequate consideration, thereby increasing their susceptibility to phishing attacks.
To address this critical challenge, research on usable security has emphasized the importance of creating polymorphic warning interfaces in phishing detection. These interfaces dynamically alter their appearance and content each time they are presented to the user, aiming to reduce habituation and encourage users to pay closer attention to each alert. By introducing variability in warnings, the polymorphic approach enhances user engagement and prompts cautious behavior, making security warnings more effective.\\
Cyri thoughtfully incorporates these research findings to enhance its phishing detection efficacy. It employs a user-centered interface that moves beyond generic warning dialogs by providing clear, contextualized explanations of potential phishing threats within the user's email text and exchanges. By offering detailed analyses and actionable advice articulated in clear and understandable language without excessive technical terminology, Cyri enhances user understanding of the potential risk causes in the email text and encourages proactive behavior in managing suspicious emails. Moreover, Cyri offers actionable advice based on the identified risks and features, providing practical recommendations on how the recipient should handle the suspicious email, such as verifying the sender's identity through alternative channels, avoiding clicking on embedded links, or reporting the email to the appropriate authorities, following best practices in phishing management~\cite{b22}.
\\
Cyri utilizes dynamic visual feedback by changing the interface's background color and icons based on the analysis results. For instance, when an email is identified as phishing, the background color shifts toward red color (see Figure~\ref{fig: examplePhishingInterface}) to indicate danger, with the intensity increasing based on the phishing likelihood percentage and the ``Feature Score'', which depends on the weights of the features found.

\begin{figure}[htbp]
  \centering
  \includegraphics[width=0.48\textwidth]{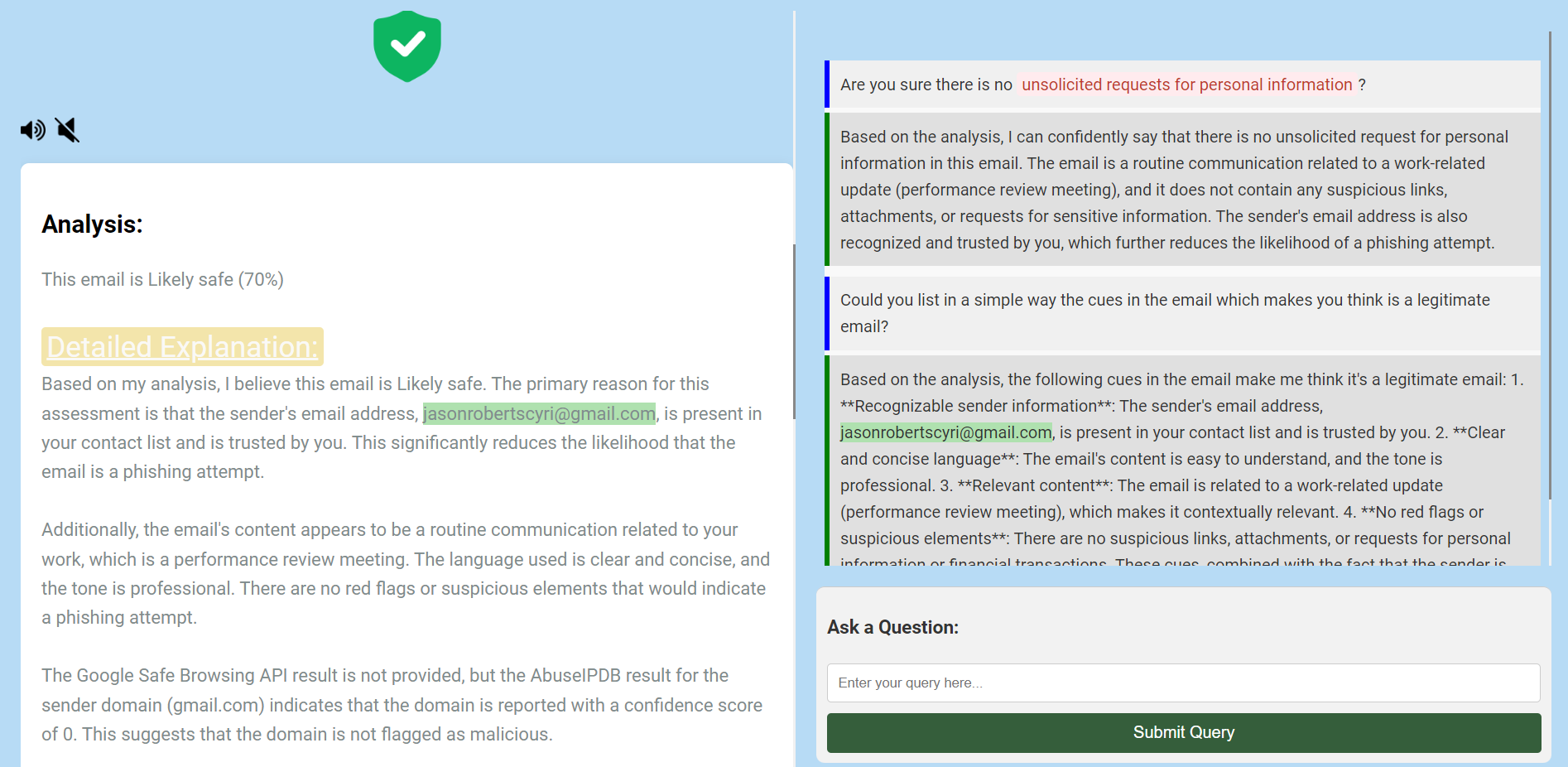}
  \caption{The VAC interface of Cyri for a safe email}
  \label{fig: exampleSafeInterface}
\end{figure}

Conversely, if an email is deemed safe (see Figure~\ref{fig: exampleSafeInterface}), the background shifts to a calming blue, with the intensity depending on the safe likelihood percentage.

\begin{figure}[htbp]
  \centering
  \includegraphics[width=0.48\textwidth]{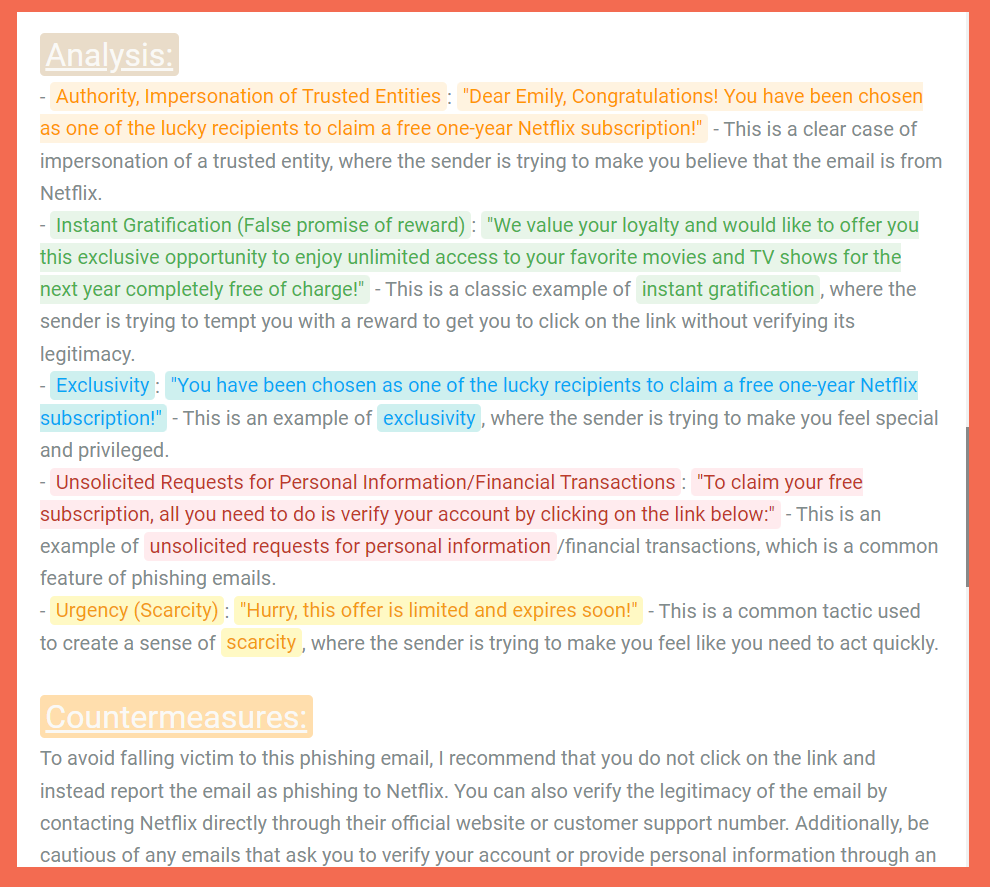}
  \caption{Phishing Email Features Analysis Example}
  \label{fig: exampleFeaturesInterface}
\end{figure}

The Cyri VAC interface is designed to offer an intuitive and user-friendly experience. The interface presents a clean and organized layout divided into two primary sections. On the left side, users encounter email details (see Figure~\ref{fig: examplePhishingInterface}.a) and analysis results (see Figure~\ref{fig: examplePhishingInterface}.b), with a date picker feature allowing effortless navigation through emails by selecting specific dates. The e-mail text keeps the same indentation and style as the email client used. Phishing features detected in the email are highlighted by utilizing unique color combinations and text styles to draw attention to these elements (see Figure~\ref{fig: exampleFeaturesInterface}).
At the bottom, there is a list of all Cyri semantic features: users can select or remove a particular feature from the analysis by clicking on it, giving them control over the information presented if they consider a specific part of the analysis wrong or not accurate enough. 

On the right side column, the application focuses on facilitating user interaction and query handling. The conversation history displays all past interactions (see Figure~\ref{fig: examplePhishingInterface}.c) and responses related to the selected email analysis, maintaining continuity and context. Users can submit new questions for additional clarification, and when a query is entered (see Figure~\ref{fig: examplePhishingInterface}.d), the application processes it and monitors for a corresponding response generated by LISA. 

The user can select tags representing phishing semantic features to navigate directly to the portions of the email body and analysis characterized by that feature (navigating them by order of occurrences or severity). 
Cyri also incorporates a text-to-speech functionality (see Figure~\ref{fig: examplePhishingInterface}.e), allowing users to have the contents of the email and the analysis results read aloud, further enhancing accessibility and allowing them to listen to Cyri's recommendations while interacting with visual cues, taking advantage of a multi-modal interaction.
\\
By utilizing dynamic and context-specific visual cues and explanations, Cyri effectively reduces habituation. Each interaction feels unique and tailored to the specific situation, maintaining the user's engagement and attentiveness to security warnings. This approach aligns with best practices in user interface design for security applications~\cite{b22, b23}, where the goal is to balance alerting users to potential threats without causing alarm fatigue.
A video demonstration of Cyri is available in the GitHub repository reported in Appendix A.

%% file: sections/06_validation.tex
\section{Validation}
\label{sec:validation}
The evaluation involved a series of tests designed to assess the model’s performance in classifying emails and identifying phishing features. The tests were structured to incrementally refine the model’s prompt and configuration presented in Section~\ref{sec:lisa} and iteratively analyze how changes affected outcomes. Key metrics such as the number of false positives, number of false negatives, Precision, Recall, and F1-score were computed to measure performance quantitatively.

\begin{itemize}
    \item \textbf{False Positives (FP)}: The number of safe emails incorrectly classified as phishing. This metric reflects instances where the model raises unnecessary alarms, causing inconvenience or mistrust.
    \item \textbf{False Negatives (FN)}: The number of phishing emails incorrectly classified as safe. This is critical as it represents missed detections, allowing potential threats to go unnoticed.
    \item \textbf{Precision}: The proportion of correctly identified phishing emails out of all emails the model classified as phishing. It measures the model's accuracy in predicting phishing emails and is calculated as: \[
    \text{Precision} = \frac{TP}{TP + FP}
    \]
    \item \textbf{Recall}: The proportion of correctly identified phishing emails out of all actual phishing emails. It assesses the model's ability to detect all phishing attempts and is calculated as:     \[
    \text{Recall} = \frac{TP}{TP + FN}.
    \]
    \item \textbf{F1-Score}: The harmonic mean of precision and recall, providing a single metric that balances false positives and false negatives. It is calculated as:     \[
    \text{F1-Score} = 2 \times \frac{\text{Precision} \times \text{Recall}}{\text{Precision} + \text{Recall}}.
    \]
\end{itemize}

\subsection{Validation Datasets}
\label{sec:valdataset}

To evaluate the model's effectiveness in detecting phishing emails, we required a comprehensive and diverse dataset of phishing emails that accurately reflect real-world phishing strategies. This dataset was meticulously crafted to encompass various topics and persuasion techniques, ensuring heterogeneity and coverage of most phishing scenarios.
\\
To create high-quality phishing emails, we utilized WormGPT V3.0~\cite{b34} since it is an advanced language model fine-tuned for crafting phishing emails (among other supported tasks). To guide the generation of phishing emails that embody the collected semantic features (both in isolation or in combination with others), we have utilized the following prompt:\\ \\
``\textit{Pretend to be a hacker planning a phishing campaign. Your goal is to generate one very detailed phishing email about this topic: \textless TOPIC\textgreater. You have to use fake American real names for the sender and recipient (example: John Smith, Sarah Thompson, Michael Garcia, Daniel Davis, James Wilson, etc) and you have to invent a phishing link URL to include in this email.''
\\ \\
The email must be characterized by the following persuasion technique:\\
\textless FEATURE \textgreater
} \\ \\
With this methodology, a total of 420 phishing emails have been generated, with 20 emails dedicated to each of the 21 identified Cyri semantic features. The emails were crafted to cover a wide array of topics relevant to each feature, enhancing the dataset’s heterogeneity. Each generated phishing email underwent a meticulous manual review by two experts to ensure that each email effectively embodied the specified semantic feature convincingly and eventually manually modified to make them more fitting to the targeted feature(s). Moreover, several additional characteristics for effective phishing attack generation, present in research papers that studied phishing characteristics, were taken into consideration such as Credibility \cite{b25}, Compatibility \cite{b25}, Personalization \cite{b35}, Contextual Relevance \cite{b35}  and Knowledge \cite{b10}, Reputation Exploitation \cite{b10}, Commitment and Consistency \cite{b20} and Liking \cite{b20}.

A meticulous process was undertaken to obtain accurate ground truth for identifying phishing features within these emails. The process involved leveraging ChatGPT-4o, supplemented by manual review and additions, to identify the specific features present in each phishing email.
\\ \\
To balance the dataset with an equal proportion of legitimate e-mails, and in the absence of an appropriate public dataset for them, a safe emails dataset composed of 420 legitimate emails was generated using ChatGPT-4. Prompts were crafted to create authentic, legitimate emails covering various topics, including:
\begin{itemize}
    \item Business Emails: Meeting requests, project updates, performance reviews, team announcements;
    \item Marketing Emails: Product launches, seasonal sales, newsletters;
    \item Personal Emails: Friendly catch-ups, event invitations, thank-you notes, congratulations messages, holiday greetings.
\end{itemize}

The two datasets were then merged into the final one, composed of 840 e-mails, heterogeneous in semantics and tactics, that will be made freely available as a public resource (the generation and curation process allow this step without incurring loss of privacy issues).

\subsection{Validation Results}
\label{sec:quantitativevalidation}

\subsubsection{Validating LLM choice}
In our comprehensive evaluation, we systematically tested the performance of the LLaMA 3.1 8B model to confirm its usage inside the LISA component. We used progressively refined prompts to assess their ability to detect phishing emails accurately.  \\ \\
In the initial phase of our evaluation, we utilized a straightforward prompt that asked whether an email was phishing or safe without providing any semantic features or detailed descriptions to guide the model's reasoning. This test aimed to establish a baseline for the model's inherent ability to classify emails based solely on its pre-trained knowledge and without additional context. The model's performance in this baseline test revealed moderate limitations (see Table~\ref{tab:test1}). There were 80 false positives, where legitimate safe emails were incorrectly classified as phishing, and 70 false negatives, where phishing emails were mistakenly identified as safe. This indicates that the model struggled to accurately differentiate between phishing and safe emails without explicit guidance. No evaluation of semantic features was possible in this case, as the problem was formulated as one of binary classification. No matter what, it confirmed our choice, dictated by security and privacy reasons, that even a small model like LLaMA 3.1 8B was a good base to build on our approach.

\begin{table}[ht]
\centering
\caption{Test 1: Classification Performance Metrics}
\label{tab:test1}
\begin{tabular}{lcccc}
\toprule
& \textbf{Precision} & \textbf{Recall} & \textbf{F1-score} & \textbf{Support} \\
\midrule
\textbf{Safe}       & 0.83 & 0.81 & 0.82 & 420 \\
\textbf{Phishing}   & 0.81 & 0.83 & 0.82 & 420 \\
\midrule
\textbf{Accuracy}   & & & 0.82 & 840 \\
\textbf{Macro Avg}  & 0.82 & 0.82 & 0.82 & 840 \\
\textbf{Weighted Avg} & 0.82 & 0.82 & 0.82 & 840 \\
\bottomrule
\end{tabular}
\end{table}

\subsubsection{Validating LISA phishing detection}
\label{sec:vallisa}

In the second test, we tried to enhance the model’s performance by incorporating semantic features of phishing emails into the prompt (see Section~\ref{sec:lisa} for details).
Results are visible in Table~\ref{tab:test2}. The number of false positives increased to 33,3\% (140), indicating that more safe emails were incorrectly classified as phishing. Conversely, the false negatives decreased to 9,5\% (40), showing an improvement in the model's ability to detect phishing emails. By introducing semantic features, the model's phishing recall improved slightly. However, this improvement in recall came at the expense of phishing precision, as evidenced by the increase in false positives. The model began over-identifying phishing characteristics in safe emails, leading to more legitimate emails being incorrectly flagged. This trade-off indicates that while the model became more sensitive to phishing indicators, it lacked the ability to adequately distinguish these features in the context of safe emails, highlighting the need for a more balanced approach.

\begin{table}[ht]
\centering
\caption{Test 2: Classification Performance Metrics}
\label{tab:test2}
\begin{tabular}{lcccc}
\toprule
& \textbf{Precision} & \textbf{Recall} & \textbf{F1-score} & \textbf{Support} \\
\midrule
\textbf{Safe}       & 0.875 & 0.667 & 0.757 & 420 \\
\textbf{Phishing}   & 0.731 & 0.905 & 0.809 & 420 \\
\midrule
\textbf{Accuracy}   & & & 0.786 & 840 \\
\textbf{Macro Avg}  & 0.803 & 0.786 & 0.783 & 840 \\
\textbf{Weighted Avg} & 0.803 & 0.786 & 0.783 & 840 \\
\bottomrule
\end{tabular}
\end{table}

Building on the previous tests, the third evaluation introduced weighted semantic features to the prompt. We assigned initial weights to each feature to reflect their importance in identifying phishing emails. Additionally, we included definitions of both phishing and safe emails and provided one example of each to guide the model exploiting 1-shot learning into the existing Chain-of-Tought approach. 
This test evidenced very good results in terms of recall (see Table~\ref{tab:test3}), with only two false negatives, indicating it almost perfectly identified all phishing emails. However, the false positives increased substantially to 42,9\% (180), resulting in many safe emails being incorrectly classified as phishing. The significant drop in precision and the high number of false positives indicated a problem with overfitting to the phishing characteristics. This imbalance suggested that the weighting scheme needed refinement to improve precision without compromising recall. 

\begin{table}[ht]
\centering
\caption{Test 3: Classification Performance Metrics}
\label{tab:test3}
\begin{tabular}{lcccc}
\toprule
& \textbf{Precision} & \textbf{Recall} & \textbf{F1-score} & \textbf{Support} \\
\midrule
\textbf{Safe}       & 0.992 & 0.571 & 0.725 & 420 \\
\textbf{Phishing}   & 0.699 & 0.995 & 0.822 & 420 \\
\midrule
\textbf{Accuracy}   & & & 0.783 & 840 \\
\textbf{Macro Avg}  & 0.846 & 0.783 & 0.773 & 840 \\
\textbf{Weighted Avg} & 0.846 & 0.783 & 0.773 & 840 \\
\bottomrule
\end{tabular}
\end{table}

In the final evaluation, we implemented a comprehensive and optimized prompt corresponding to the one utilized by the Cyri system. 
To address the issues identified in the previous test, we adjusted the weights assigned to the features, aligning them more appropriately with their actual importance in phishing detection. We also provided multiple examples (3-shot learning) of safe emails to enhance the model's understanding of legitimate email patterns, having identified FP as the most problematic case. Finally, we enhanced the prompt to encourage the model to utilize its inherent reasoning abilities alongside all the detailed information provided, following public heuristics on how to make a prompt-based strategy more effective. This approach allowed the model to perform a more comprehensive analysis by ensuring that the model’s general understanding and language comprehension are utilized initially, potentially capturing nuances that this particular feature-based analysis might miss. 
These changes produced significant improvements in performance metrics. There were 13 false positives and 27 false negatives.

\begin{table}[ht]
\centering
\caption{Test 4: Classification Performance Metrics}
\label{tab:updated-classification-metrics}
\begin{tabular}{lcccc}
\toprule
& \textbf{Precision} & \textbf{Recall} & \textbf{F1-score} & \textbf{Support} \\
\midrule
\textbf{Safe}       & 0.938 & 0.969 & 0.953 & 420 \\
\textbf{Phishing}   & 0.968 & 0.936 & 0.952 & 420 \\
\midrule
\textbf{Accuracy}   & & & 0.952 & 840 \\
\textbf{Macro Avg}  & 0.953 & 0.952 & 0.952 & 840 \\
\textbf{Weighted Avg} & 0.953 & 0.952 & 0.952 & 840 \\
\bottomrule
\end{tabular}
\end{table} 

\subsubsection{Validating LISA phishing semantic features detection}
\label{sec:valsemfeat}

\begin{figure}[htbp]
  \centering
  \includegraphics[width=0.48\textwidth]{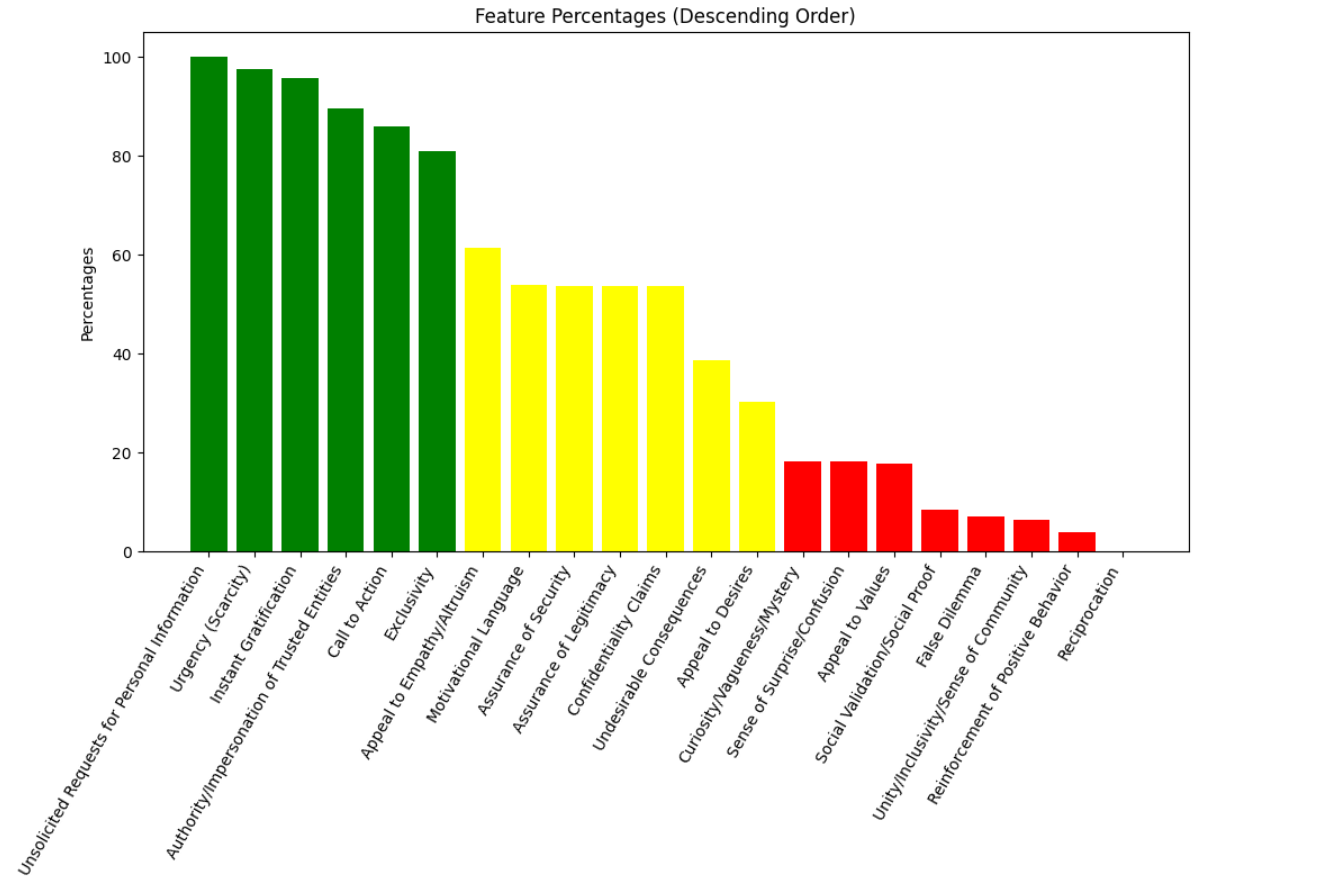}
  \caption{Final configuration for semantic features detection accuracy}
  \label{fig:testFeatures}
\end{figure}

An in-depth validation was also conducted to evaluate the LISA's ability to identify specific phishing features by comparing the list of features found by LISA with the ground-truth list of features previously curated using ChatGPT-4o and manual annotations. The features were categorized based on the percentage of correct identifications into three classes of accuracy and sorted by decreasing accuracy. These results are visible in Figure~\ref{fig:testFeatures}.

High accuracy rates were observed for critical semantic features fundamental to phishing detection, such as Unsolicited Requests for Personal Information/Financial Transactions, Urgency (Scarcity), Authority/Impersonation of Trusted Entities, Call to Action, and Exclusivity.
In those cases, we rarely foresee doubts from human users (expert or non-expert) about the presence of these features in a suspicious e-mail, but taking advantage of their identification as a severe factor for not trusting the message.
\\
Medium accuracy rates were noted for features like Appeal to Empathy/Altruism, Motivational Language, Assurance of Security, Undesirable Consequences, Curiosity/Vagueness/Mystery, and Sense of Surprise/Confusion.
For these cases, severity is lower, and contextual factors may help a non-expert user assess the genuine or malicious nature of the messages. No matter what, Cyri alerts them, helping the human user confirm or deny their harmful nature through visual inspection and conversation.
\\
Features with lower accuracy rates included Appeal to Values, Social Validation/Social Proof, False Dilemma, Reinforcement of Positive Behavior, and Reciprocation.
We found this result coherent with a situation where these characteristics may also happen in genuine email or more subtle tentative. These areas need to be improved from an automatic detection point of view, and this represents a current limitation of our approach that needs further investigation. Possible mitigations may be providing visual alerts in the absence of these characteristics or providing for these characteristics the information on the degree of confidence of LISA for consideration and further analysis. 

\subsubsection{Overall results discussion}
\label{sec:overallvalidation}

The final model demonstrated a significant improvement in both precision and recall, achieving a high level of accuracy. The balanced weighting of features, comprehensive definitions, and multiple examples contributed to reducing both false positives and false negatives. The feature identification analysis revealed that the model was highly effective in detecting critical phishing features. High accuracy rates for key features affirm the model's effectiveness in accurately identifying phishing emails. Features with lower accuracy rates were less crucial for phishing detection, and their misidentification did not substantially impact the overall performance. However, these features might benefit from well-structured fine-tuning to enhance the model's comprehensiveness and explanatory capabilities.

%% file: sections/07_evaluation.tex
\section{User Evaluation}
\label{sec:eval}
To assess the effectiveness and usability of Cyri as a tool for phishing detection and management from a human user, a user study was conducted involving ten participants with varying levels of expertise in computer security. This section details the methodology of the user study, the setup, and discusses the findings.

\subsection{Experiment Setup}
\label{sec:setup}

The study involved 10 participants, split equally between computer security
experts (meaning having at least two years of expertise and being knowledgeable of phishing tactics and techniques) and non-experts (meaning not being knowledgeable of phishing tactics and techniques but capable of using an email account).\\
The study duration for each participant was 60 minutes, split into 15 minutes of initial explanation on what are the most important features of Cyri and how to install it. This first step was then followed by two main tasks:
\begin{itemize}
    \item Controlled Email Identification Task: Participants were put in front of a preconfigured installation of Cyri and received five emails, four safe emails, and one phishing email sent by us. They were instructed to review these emails with Cyri and identify the phishing emails among them and the motivating factors for their final decision.
    This test was used both to let the participants gain confidence with Cyri usage and interface and as a controlled experiment where to evaluate how users interpreted and used the different results and functionalities Cyri exposes in a controlled situation equals for all of them. This step lasted, on average, from 10 to 15 minutes.
    \item Exploration with Personal Emails: Participants were then tasked to use Cyri to analyze their inbox emails from one personal account, such as those in their spam folder, unopened ones, or newly received messages. This allowed them to interact with the application in a context familiar to them and to assess its usefulness beyond the controlled task of provided emails, resulting in a more personal experience capable of letting them assess the degree of support they received from Cyri. This task lasted, on average, 25 minutes.
\end{itemize}
After completing the second task, participants were asked to compile a survey
comprising several questions aimed at evaluating Cyri’s effectiveness in assisting users in identifying phishing emails, usability and intuitiveness of the application interface, impact on users' understanding of phishing tactics, the likelihood of continued use, and preference for platform availability.
In particular, the questions proposed to the participants were the following:

\begin{enumerate}
    \item Are you a computer security expert? (Yes or No)
    \item How confident are you in identifying phishing emails without assistance?  (Scale 1 to 5)
    \item How useful was Cyri in helping you identify the phishing email?  (Scale 1 to 5)
    \item Did Cyri provide information that you wouldn't have noticed on your own? (Yes or No)
    \item How would you rate the overall usability of Cyri? (Scale 1 to 5)
    \item How intuitive did you find the Cyri interface?  (Scale 1 to 5)
    \item How much do you think using Cyri would improve your understanding of phishing tactics? (Scale 1 to 5)
    \item Would you use Cyri regularly as part of your email routine?  (Yes or No)
    \item Would you prefer if Cyri was available on your mobile phone instead of your computer?  (Yes or No)
\end{enumerate}

A final free text form allows the insertion of open comments and suggestions. Overall 5 minutes were dedicated on average to this activity.

\subsection{Results}
\label{sec:userresults}

\begin{figure}[htbp]
  \centering
  \includegraphics[width=0.45\textwidth]{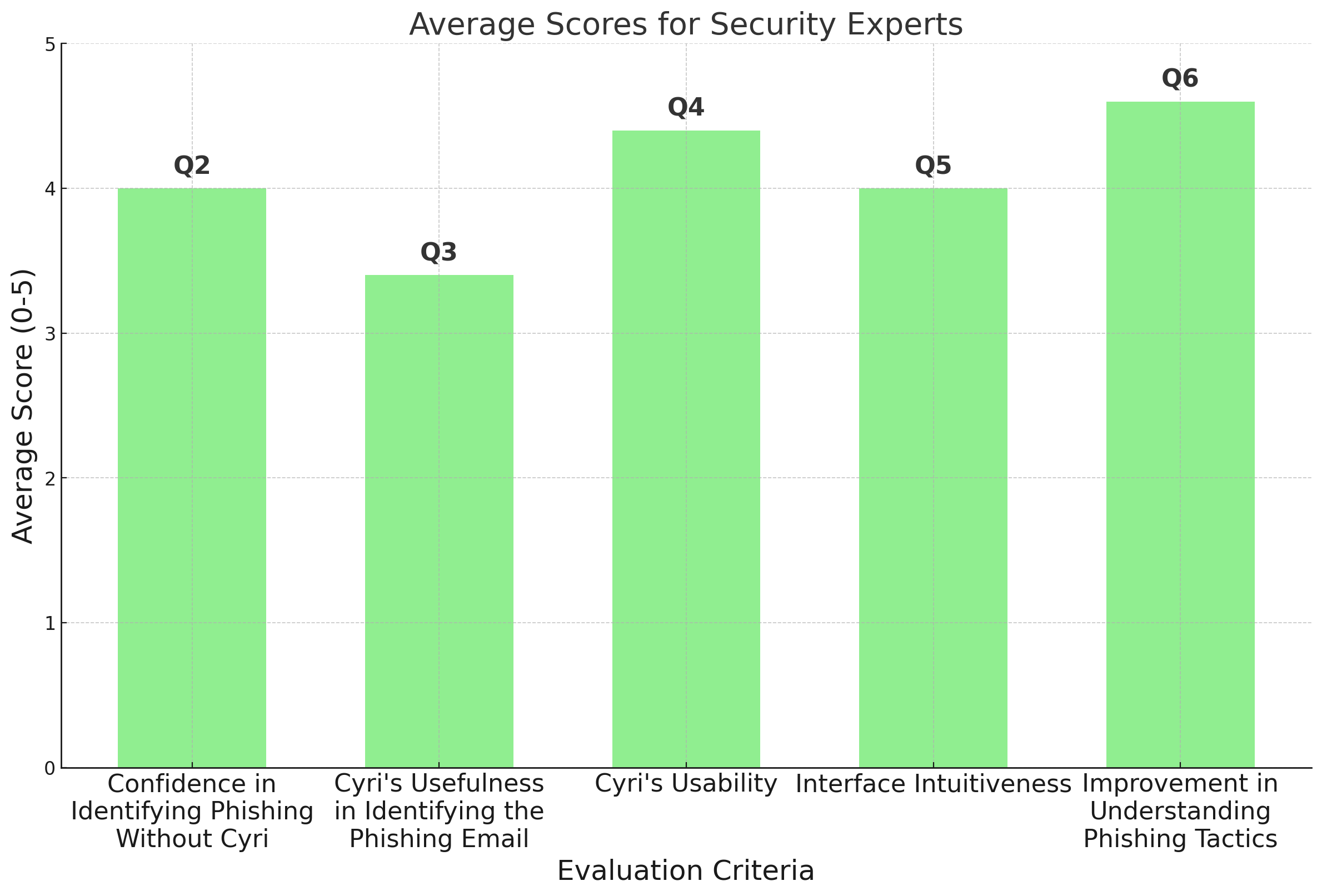}
  \caption{Security Experts Average Scores}
  \label{fig:secexp}
\end{figure}

\begin{figure}[htbp]
  \centering
  \includegraphics[width=0.45\textwidth]{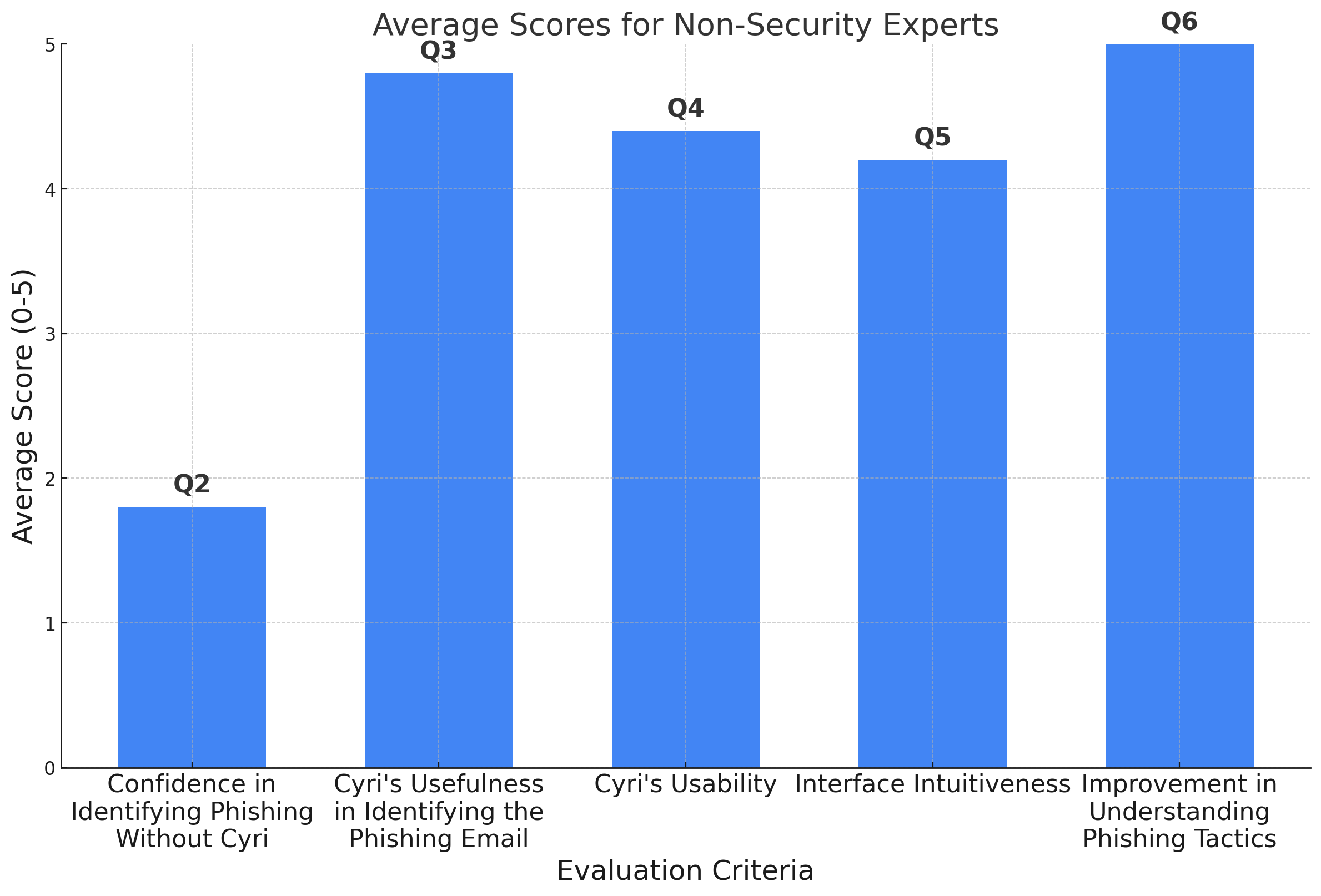}
  \caption{Non-Security Experts Average Scores}
  \label{fig:nonsecexp}
\end{figure}

We analyzed the survey results by splitting participants into their expertise level into two groups: Figure~\ref{fig:secexp} reports results for computer security experts while Figure~\ref{fig:nonsecexp} reports them for non-expert users. This distinction allowed us to understand how Cyri is perceived by users with different levels of expertise.

Cyri has been declared to be highly beneficial by non-expert participants, reporting generally low confidence in their ability to identify phishing emails without Cyri's assistance (Q2). All non-expert participants affirmed that Cyri provided information they would not have noticed on their own (Q4). This suggests that Cyri effectively highlights phishing indicators that might be overlooked, adding significant value in assisting them in identifying potential threats. 
Furthermore, non-experts provided very high ratings for both the usability (Q5) and intuitiveness (Q6) of Cyri. These results indicate that they found the application user-friendly and accessible. Non-experts believed using Cyri would significantly improve their understanding of phishing tactics (Q7), underscoring the application’s educational value.

Expert participants, even if they declared an expected good capability of identification and management of the phishing email with and without Cyri support (Q2), acknowledged that Cyri can enhance analysis capabilities by providing an additional layer of information (Q3). Interestingly, all expert participants also affirmed that Cyri provided information they would not have noticed on their own (Q4). This indicates that Cyri can uncover subtle phishing indicators and offer insights that even experienced users might overlook. Experts rated the overall usability (Q5) and intuitiveness (Q6) of Cyri highly, similar to non-experts, suggesting that the application is well-designed for users across different expertise levels. Moreover, experts believed that using Cyri could further improve their understanding of phishing tactics (Q7).
All participants expressed their willingness to use Cyri regularly as part of their email routine (Q8) and showed a clear preference for having it available also on their mobile devices (Q9).

%% file: sections/08_discussion.tex
\section{Discussion}
\label{sec:discussion}
While Cyri enhances phishing detection, management, and understanding from human users, its current implementation also presents a set of limitations that offer avenues for further improvement.
Improvements in detecting low-accuracy semantic features could be achieved by fine-tuning activities leveraging the produced phishing email dataset. These features, while less critical than others, still contribute to the overall understanding of phishing tactics. This solution would not be in substitution but would complement the current Chain-of-thought approach. Retrieval Augmented Generation can even be exploited, using currently detected emails as additional context for more accurate detection.
Another limitation we foresee is the need for a longitudinal study with users that lasts longer and collects usage data on a higher quantity of tested emails and in real-pressure conditions. We are planning this activity in the near future.


As interesting future possibilities enabled by this research, we foresee integrating Cyri into existing mobile email clients, which would enhance accessibility and provide real-time phishing detection and education on the devices most commonly used for email communication. Limits and possibilities in this scenario could be provided by quantized versions of small LLMs capable of being run on smartphones with similar accuracy to 8 billion models. The use of information distillation techniques with a teacher-student approach using LISA as the teacher model may be beneficial for this effort.


%% file: sections/09_conclusions.tex
\section{Conclusions}
\label{sec:conclusions}

This paper contributes Cyri, a significant advancement in applying AI-driven
solutions for phishing detection and management for human users. 
Through a systematic collection of semantic features and the tuning of a local LLM to detect them directly in email text, we created a system capable of detecting subtle cues indicative of malicious intent that traditional methods might overlook.
Extensive iterative testing and prompt refinement were conducted to optimize
Cyri's performance. By providing detailed explanations and engaging in conversational interactions through its visual and conversational interface, Cyri helps users understand why an email is potentially malicious and what steps they should take.
This addresses the human factors contributing to phishing success, such as lack of awareness and susceptibility to psychological manipulation.
Cyri achieved good results in detecting and explaining phishing emails and very positive results for efficacy and usability by ten experts and non-expert human users in a task-based evaluation.
In future work, we plan to mitigate the reported limitations of Cyri and explore its use of Cyri not only as a detection tool but also as an educational tool for training and awareness activities.